\documentclass[11pt,preprint]{aastex}
\def\msun{\rm M$_{\sun}$}

\def\lsun{\rm L_{\sun}}
\def\teff{{$\rm{T_{eff}}$ }}

\begin{document}
\shortauthors{Hern\'andez et al.}
\shorttitle{HAeBe in associations}
\title{Herbig Ae/Be Stars in nearby OB associations }

\author{Jes\'us Hern\'andez\altaffilmark{1,2,3},
Nuria Calvet\altaffilmark{4,2}, 
Lee Hartmann\altaffilmark{4},
C\'esar Brice\~no\altaffilmark{1,2}, 
Aurora Sicilia-Aguilar\altaffilmark{4},
and Perry Berlind\altaffilmark{4}
}
\altaffiltext{1}{Centro de Investigaciones de Astronom{\'\i}a (CIDA),
Apartado Postal 264, M\'erida 5101-A, Venezuela;
Electronic mail: jesush@cida.ve, briceno@cida.ve}
\altaffiltext{2}{Postgrado de F{\'\i}sica Fundamental, Universidad de Los Andes (ULA),
M\'erida 5101-A, Venezuela}
\altaffiltext{3}{Visiting Student, Harvard-Smithsonian Center for Astrophysics}
\altaffiltext{4}{Harvard-Smithsonian Center for Astrophysics, 60
Cambridge, MA 02138, USA, Electronic mail:
ncalvet@cfa.harvard.edu, hartmann@cfa.harvard.edu,pberlind@cfa.harvard.edu}

\begin{abstract}
We have carried out a study of the early type
stars in nearby OB associations spanning an age range of
$\sim$ 3 to 16 Myr, with the aim of determining
the fraction of stars which belong to the
Herbig Ae/Be class.
We studied the B, A, and F stars in the
nearby ($\le 500$ pc) OB associations Upper Scorpius, Perseus OB2, Lacerta OB1,
and Orion OB1, with membership determined from
Hipparcos data.
We also included in our study the early stars in the
Trumpler 37 cluster, part of the Cep OB2 association.
We obtained spectra for 440 Hipparcos stars in these
associations, from which we determined
accurate spectral types, visual extinctions, effective temperatures,
luminosities and masses, using Hipparcos photometry.
Using colors corrected for reddening, we find that
the Herbig Ae/Be stars and the Classical Be stars (CBe)
occupy clearly different regions in the JHK diagram.
Thus, we use the location on the JHK diagram, as well as 
the presence of emission lines and of strong 12 $\mu$m
flux relative to the visual to identify the Herbig Ae/Be stars
in the associations.
We find that the Herbig Ae/Be stars
constitute a small fraction of the early
type stellar population even in the younger associations.
Comparing the data from associations with different ages and
assuming that the near-infrared excess in the
Herbig Ae/Be stars arises from optically
thick dusty inner disks, we 
determined the evolution of the inner disk
frequency with age. We find that the inner disk
frequency in the age range 3 - 10 Myr in intermediate mass stars is
lower than that in the low mass stars ($<$ 1 \msun);
in particular, it is a factor of $\sim$ 10 lower
at $\sim$ 3 Myr.
This indicates that the time-scales for disk evolution
are much shorter in the intermediate mass stars,
which could be a consequence of more efficient
mechanisms of inner disk dispersal
(viscous evolution,
dust growth and settling toward the midplane).
\end{abstract}


\section{Introduction}
\label{sec:int}

	The Herbig Ae/Be stars (HAeBe) are young emission line 
objects with spectral types B, A, and in a few cases F, in most 
instances spatially  correlated with dark clouds or bright 
nebulosities \citep{herbig60,waters98,hernandez04}.
The masses of HAeBe range from 2 to 10 {\msun}.
Like their low mass pre-main-sequence (PMS) counterparts, 
the T Tauri stars (TTS), these objects become optically 
visible before reaching the main sequence, so their
PMS evolution can be studied in some detail. 
Stars more massive than 10 {\msun} are expected to spend their whole 
PMS time as optically obscured objects.

In addition to H$\alpha$ emission, HAeBe 
exhibit infrared excesses relative to the photosphere which are 
attributed to emission from dust in circumstellar accretion disks
\citep{finkenzeller84,lorenzetti83, davies90,hill92,ancker97,malfait98}.
Millimetric and submillimetric observations confirm the existence of 
disks of substantial mass around some of these objects 
\citep{blake04,fuente03,mannings00,mannings97,natta00, natta01}.
Most HAeBe exhibit a flux excess in the near 
infrared (NIR) portion of their spectral energy distributions (SEDs), 
which has been attributed to emission
from an optically thick ``wall'' at the dust destruction radius,
directly heated by the star \citep{natta01, maheswar02,dullemond01,dullemond04a,muzerolle04, muzerolle03b}.

	Studies of the circumstellar environments of HAeBe suggest 
that  these objects are progenitors the stars surrounded by 
debris disks ( e.g, $\beta$ Pictoris, Vega-like stars), which could be 
sites of planet formation \citep{natta00, lagrange00}. 
However,  long standing questions about how and when the transition
 between HAeBe  and Vega-like stars occurs have yet to be answered. 
Observations  of  the numbers of young intermediate mass stars 
surrounded by optically thick inner disks in samples of  
different ages could be used as  a first step toward answering these 
 questions. Roughly coeval stellar groups are ideal to
 study since uncertainties in properties like distances and 
ages are minimized.

	OB associations are defined as stellar groups with significant
populations of intermediate mass stars and with stellar mass density 
less than 0.1 \msun/$pc^3$ \citep{brown99b}. 
Ages derived from Hertzsprung-Russell diagrams (HRD)
 indicate that they are young \citep[e.g,][]{blaauw64}.
For this reason, OB associations are excellent for studying young 
stellar objects covering a complete range of mass \citep[e.g. ][]{preibisch02}. 
Moreover, by studying OB associations with different ages, 
we can analyze the temporal evolution of the processes 
characterizing the early history of their stars.

	An important but difficult aspect
of studying OB associations is the identification of their members. 
OB associations have small internal velocity dispersions, 
so the streaming motion of 
the stellar group, in combination with the solar motion, 
is reflected as a motion of their members toward a convergent 
point on the sky \citep[e.g.,][]{brown99a,zeeuw99}. Thus, 
membership can be established by studying the kinematic 
properties of the group (proper motions, parallaxes, radial 
velocities). Nearby associations cover large 
regions in the sky, which in the past limited the astrometric 
membership determination to the bright stars (V $\leq$ 6 mag). 
Photometric studies could add fainter members to the 
associations, but this membership determinations were less reliable due
to several factors as, for instance, undetected duplicity, spread of distances 
within the association, photometric variability, 
and interlopers. The publication 
of positions, proper motions and trigonometric parallaxes 
for $\sim$120000 stars in the Hipparcos Catalog \citep{esa97} 
has significantly improved the situation, 
enabling astrometric studies 
down to its limiting magnitude, V$\sim$12. 

	Using  Hipparcos measurements, \citet{zeeuw99} made a 
comprehensive census of the stellar content of the OB 
associations within 1 Kpc from the sun. Members 
of each association were identified using two methods.
The first method is a modification of the classical convergent point 
method \citep{brown50} detailed by \citet{bruijne99};
the second uses the positions, parallaxes and 
proper motions to define members using probability 
distributions in the velocity space \citep{hooger99}. The combination of
these methods gives reliable memberships of stars in 
12 young stellar groups. However, since some groups are beyond 500 pc,
where the Hipparcos parallaxes are no longer useful, or have unfavorable
kinematics, 10 associations studied by \citet{zeeuw99} do not show
astrometric evidence for a moving group and their results
are inconclusive. One of these associations 
with unfavorable kinematics is Orion OB1, which has a motion mostly
directed radially away from the Sun.
Nonetheless, the Hipparcos 
data of Ori OB1 was analyzed carefully by \citet{brown99b}, 
who found a relation based on proper motions which characterizes
approximately the members of this association (see \S 2.2). 
The resulting set of stars
selected by this relation overlaps in 96\% with the photometric 
members given by \citet{brown94}.

	In this contribution we explore the frequency of HAeBe in 
three associations with different ages (Upper Scorpius, 
Perseus OB2, and Lacerta OB1) with members selected from \citet{zeeuw99}. 
In addition, we study the Orion OB1 association,  which was divided 
in 2 sub-association with different ages and distances. 
We complement our studies by analyzing the NIR properties of 
the early stars in Trumpler 37  from \citet{contreras02}. 
Details of the selection and the observation of each sample is 
described in \S2. In our entire sample, more than 82\% of the 
spectral types published by Hipparcos 
are from the SIMBAD\footnote{SIMBAD, Centre de Donn\'{e}es 
astronomiques de Strasbourg: http://simbad.u-strasbg.fr/sim-fid.pl} 
data-base, with references labeled as miscellaneous; the other
18\% are from the Michigan catalogue for the HD stars \citep{houk82,houk88}. 
We obtained spectra for all the stars in our sample and
reclassified them using the classification scheme described in
\citet{hernandez04},  to obtain an homogeneous spectral 
typing scheme for all the objects in the associations. Details of 
the classification scheme are in \S 3.1, where we also calculate 
the visual extinction for the objects and describe
the identification and measurement of the emission lines present in the 
spectra.  We analyze the NIR excess of the members of each group in \S 3.2. 
We estimate distances from both astrometric and photometric data in \S 3.3, 
and a HRD  is given for each association in \S 3.4. 
A census of HAeBe, based on observational properties 
characteristic of the presence of disks (emission at H$\alpha$, 
NIR excesses and IRAS fluxes) is described in \S 4, and the
disk frequency is discussed on \S5. Finally, we summarize the most 
important aspects of this work in \S6.

\section{Selection of the sample and observations}
\label{sec:sample}

	Table 1 lists properties of the associations studied in 
this contribution. Our total sample include 440 Hipparcos stars distributed 
in 3 OB associations (\S2.1), with astrometric membership determined by 
\citet{zeeuw99}, and the Orion OB1 association, for which we have determined 
membership combining astrometric and photometric data (\S 2.2).

\subsection{Upper Scorpius, Perseus OB2 and Lacerta OB1 sample}

       Using the census of nearby  OB associations made
by \citet{zeeuw99}, we selected young associations 
with estimated ages $<$ 20 Myr and within 500 pc from the sun, since
the uncertainty in the Hipparcos parallaxes beyond 
500 pc is generally larger than 50\%. We also required  the 
associations to be within the declination limit of the telescope 
used in this work ($\delta >-35 ^\circ$, see \S 2.3).  With this 
criteria we selected the OB associations Upper Scorpius (US), 
Perseus OB2 (Per OB2), and Lacerta OB1 (Lac OB1).  
Columns 2, 3, 4 of Table 1 give the distances, the ages, 
and the number of stars for each association, respectively 
\citep{zeeuw99}. Column 5 of this table gives 
the number of stars observed for each one of these three associations. 
Other information in this table is discussed below.

\subsection{Orion OB1 sample}

	As we said previously, the association Orion OB1 has unfavorable
kinematics, so the membership determination method from
\citet{zeeuw99}  is not useful. However, \citet{brown99b} and 
\citet{brown94} studied in detail the brighter population of this association 
giving several criteria to find members belonging to Ori OB1. These
criteria are applied in this section to create our sample.

     We defined a region of 180 square degrees in Orion OB1 
defined by limits in right ascension $5.0h <\alpha < 6.0 h$  
and declination $ -6^\circ <\delta < 6^\circ $. 
There are 733 Hipparcos stars in this region. 
Since the motion of the Ori OB1 association is mostly directed 
radially away from the Sun, the expected intrinsic proper
motions in right ascension and declination, $\mu_\alpha$ and
$\mu_\delta$, 
have to be small and comparable to measurement errors. 
So, we rejected stars with a relative large intrinsic proper motion 
by applying  the criteria defined by \citet{brown99b},

\begin{equation} 
(\mu_{\alpha} cos(\delta) - 0.44)^2 + (\mu_{\delta}+0.65)^2 \leq 25        
\end{equation}

Where the proper motions are in milli-arcsecond per year (mas/yr)
 
	This method provides a rough selection of members in Orion OB1.
From the 282 stars selected by this criterion, we rejected 
14 stars because they have negative parallaxes. Another 4
stars were rejected because they are clearly foreground stars with parallaxes 
larger than 7 mas and corresponding to distances smaller than 140 pc. 
In addition, using photometry from the Hipparcos catalog,
we applied a photometric criterion
 in the B-V vs V color-magnitude diagram, selecting
 stars located above or on the zero age main sequence
\citep[ZAMS;][]{allen2000} and with B-V color < 1.2 in order 
to avoid highly embedded objects or intrinsically red stars 
(K and M). 
The final set of candidates include 245 objects,
of which 92\%  were observed spectroscopically (see Table 1).  

	Four subgroups of Ori OB1 with differences in ages and 
distances (1a, 1b, 1c and 1d) were identified by \citet{blaauw64}. 
These  subgroups were largely analyzed in later studies 
\citep[e.g.,][]{warren77a, warren77b, genzel89, brown94}. 
Particularly, \citet{brown94} determined ages 
(1a: 11.4 $\pm$ 1.9 Myr; 1b: 1.7 $\pm$ 1.1 Myr; 1c: 4.6 $\pm$ 2.0 Myr) 
and distances ( 1a: 380 $\pm$ 90 pc, 1b: 360 $\pm$ 70 pc; 
1c: 400 $\pm$ 90 pc) based on the photometric properties 
of each subgroup. \citet{brown99b} improved the estimate of 
distances  based on Hipparcos parallaxes ( 1a: 336 $\pm$ 16 pc, 
1b: 439 $\pm$ 33 pc; 1c: 462 $\pm$ 36 pc). The distance of the youngest
subgroup Ori OB1d ($<$ 1Myr), located at the Orion Nebula Cluster (ONC), was 
undetermined due to the small number of stars in the \citet{brown99b} study.

Figure 1 shows stars in the Orion OB1 association overlaid
on a map of integrated  $^{13}$CO emissivity
from \cite{bally87} (gray scale) and isocontours of
galactic extinction  for A$_V$ = 1, 2, 3 and 4, from \cite{schlegel98}.
We plot the boundaries from \citet{warren77a} between the subgroups 1a, 
1b, and 1c.
Due to the similarities in the ages and distances of the sub-associations 
OB1b and OB1c, we have joined these sub-associations 
into one (labeled as OB1bc). 
Most of the stars in OB1bc are spatially related with dust or gas. 
There is a subset of 17 stars in the OB1a sub-association which is
also spatially correlated with dust and gas (see Figure 1);
we assumed that this subset belongs to OB1bc instead of OB1a. 
A Kolmogorov-Smirnov 
test based on Hipparcos parallaxes supports this assumption. 
This test shows that the significance level is 10\% higher  
when the parallaxes of this subset are compared to the parallaxes of OB1b 
than when they are compared to the parallaxes of the remainder OB1a stars. 
The last two rows in Table 1 give  
information about the subgroups in Ori OB1.

\subsection{Observations}
We obtained low-dispersion spectra for the stars selected  
as discussed in \S 2.1 and \S 2.2 during 2000 and 2003 
using the 1.5 meter telescope of the  Whipple Observatory with the
FAST Spectrograph \citep{fabri98}, equipped with the Loral 
$512 \times 2688$ CCD. The spectrograph was set up in the standard configuration 
used for ``FAST COMBO'' projects, a 300 groove $mm^{-1}$ grating and a 3'' 
wide slit. This combination offers 3400 {\AA } of spectral coverage centered
at 5500 {\AA}, with a resolution of 6 {\AA }. 
The spectra were reduced at the CfA using software developed specifically 
for FAST COMBO observations. All individual spectra were wavelength 
calibrated and combined using standard IRAF routines
\footnote{IRAF is distributed by the National Optical Astronomy Observatories,
which are operated by the Association of Universities for Research
in Astronomy, Inc., under cooperative agreement with the National
Science Foundation.}. 
The exposure time range from 0.2 to 180 seconds. 
Signal-to-noise ratio (SNR) of our spectra are typically $\ga 20$ 
at the central wavelength region of the spectra. 
The standard stars used in our spectra classification scheme 
were observed with the same  instrumental configuration 
used for the Hipparcos stars.

\section{Analysis of the observations}

Table 2 compiles properties derived from our observations.
Column 1 and 2 show the Hipparcos number and other name of the stars. 
Spectral types and reddening (A$_V$), with their respective errors, 
are in columns 3, 4, 5, and 6 (see \S 3.1). The 2MASS colors 
used in \S 3.2 are shown in columns 7, 8 and 9. The effective 
temperature (\teff), luminosity and mass
calculated in \S 3.4 are given in the last three columns.

\subsection{ Spectral analysis and reddening estimates }

 We classified ours objects following the spectral classification 
scheme of \citet{hernandez04}, which is optimized for the FAST wavelength range 
and is defined for stars with spectral types in the range from O to early G stars. 
The scheme is based on 33 spectral indices sensitive to changes 
in \teff but insensitive to reddening, stellar rotation, luminosity class and 
signal-to-noise ratio. 
This scheme is designed to largely avoid problems caused by non-photospheric 
contributions (mostly due to  material surrounding the star). We achieve 
this by requiring that the various spectral types calculated from each 
index agree with the others; wildly discrepant values are rejected, 
and a weighted mean spectral type is obtained. In columns 3 and 4 of Table 2, 
the spectral type and its respective error is show for each star. This
error has two contributions, the error from the fit of each index to the
standard main sequence, and the error in the measurement of each 
index \citep{hernandez04}

We measured the equivalent width (EW$_\lambda$) of H$\alpha$ 
and of the most prominent emission lines using the task $splot$ in
the IRAF spectra reduction package $noao.onedspec$. 
In those lines with emission and absorption components, 
the EW$_\lambda$ is referred 
to the emission contribution. These emission line stars provide us with a
list of  HAeBe candidates to which we apply additional constraints, 
like NIR (\S3.2) and IRAS fluxes (\S 4.1) to select stars with 
inner dusty disks . 

We use B-V colors from the Hipparcos Catalog
In order to estimate the visual extinction A$_V$.
Only the star HIP17561 in Perseus OB1 
association does not have photometric information, 
so we take the V and B-V values from the SIMBAD data-base. The
intrinsic (B-V)$_0$ color is obtained interpolating our 
spectral type in the table of colors 
for main sequence stars given by \citet[][ hereafter KH95]{kh95}.
\defcitealias{kh95}{KH95}
We use the extinction
relation from \citet{ccm89} and the value of 
total-to-selective extinction for normal interstellar reddening
(R$_V$=3.1). Although in moderately and highly embedded HAeBe this value
is larger  than that for normal interstellar 
reddening \citep[e.g.][]{hernandez04,whittet01, waters98}, 
we use R$_V$=3.1 for our stars because they have low visual extinction 
(A$_V$ $<$ 1 mag); differences in  R$_V$ imply differences in A$_V$ 
(column 5 of Table 2) smaller than the error (column 6 of Table 2) 
propagated from errors in the spectral type and the photometric data.  
 
\subsection{Near Infrared Color Color Diagram}

The HAeBe show significant excesses in the JHK diagram
relative to their photospheric colors \citep{hill92,lada92}.
These excesses are associated with disk emission \citep{natta01, maheswar02,dullemond01,dullemond04a,muzerolle04}.
However, there are early type stars with emission lines, the CBe,
which also show near infrared excesses from gaseous free-free emission, although they are not of PMS nature;
these stars are often confused with HAeBe.
To determine the actual regions in the JHK color-color 
diagram occupied by the HAeBe and the CBe,
we have compared dereddened colors of samples of both types.
The HAeBe sample is taken from \citet{hernandez04}
and the sample of CBe from \citet{yudin01}.
The HAeBe, the CBe and the objects in 
each association were corrected for reddening 
using the relations from \citet{ccm89}. 
The dereddened 2MASS colors were converted 
into the standard CIT systems using the transformations from \citet{carpenter01}. 
The intrinsic colors of dwarf and giant stars in Johnson-Glass system 
were taken from \citet{bessell88}. 
We complete the earlier main sequence stars with photometric data from 
\citet{koornneef83} transformed to the Johnson-Glass system \citep{allen2000}. 
The Johnson-Glass system system was converted to the CIT system using the transformation 
given in \citet{bessell88}.

	In Figure 2, we plot dereddened colors  of the sample of 
HAeBe \citep{hernandez04} on the JHK color-color diagram in three 
spectral ranges, earlier than B5, later than F0, 
and between B5 and F0. 
We selected stars with  reliable 2MASS data as determined from
their photometric quality flag. We complete the NIR data using
photometry from \citet{eiroa01} and by \citet{dewinter01}.
 All these objects were corrected by reddening using 
 the A$_V$ calculated by \citet{hernandez04} and  the mean value of R$_V$
 adequate for this sample (R$_V$=5.0). 
We  also plot in Figure 2 the dereddened colors of a sample of CBe 
from \citet{yudin01}, 
distinguishing stars
with spectral type B5 or later, and
spectral type earlier than B5. 
To get a 
homogeneous CBe sample, we selected objects with 2MASS 
photometry and with HIPPARCOS spectral type taken from 
the Michigan Catalogs \citep{houk75,houk78,houk82,houk88}. 
We used the B-V color from the Hipparcos 
catalog and the normal interstellar reddening law 
to estimate the A$_V$ for the CBe.

The CBe and the HAeBe occupy separate 
regions in the JHK color-color diagram, with  
the CBe located in a relative small 
region near the blue end of the main sequence. 
In contrast, most of the HAeBe
are distributed below the Classical T Tauri star (CTTS) 
locus defined  by \citet{meyer97}, in a more extended band 
more or less parallel to the reddening vector, but displaced 
to the right of the reddening line from a B0 star (Figure 2).
This clear difference on the JHK color-color diagram 
was noted previously by \citet{lada92} 
using samples of HAeBe and Be stars, but the objects 
were not corrected for reddening, so the region occupied by
 the HAeBe in that paper was more spread out extending even 
 above the CTTS locus. 
In Figure 2, there are two stars listed as HAeBe that appear 
on the CBe region. One of them is MC 1 (HBC324) which 
has spectral type A7 and emission at the forbidden lines
[\ion{O}{1}] $\lambda$6300, [\ion{O}{1}] $\lambda$6363, 
[\ion{S}{2}] $\lambda$6717 and [\ion{S}{2}] $\lambda$6731 
additional to the emission at  H$\alpha$.
The other star, BD+651637 (HBC730), has spectral type B4 with emission at 
H$\alpha$ and H$\beta$ and \ion{Fe}{2}. These objects could be 
stars with nearly pole-on disks, where the contribution of the vertical wall
at the dust destruction radius is negligible \citep{dullemond01}; 
however, no IRAS sources are associated with them.
Additional studies of these stars are required to clarify if the 
lack of NIR excess is produced by a special geometry of the disk or by
its absence. The object located above of the CTTS locus 
is the star V633 Cas (HBC3); the strong IRAS color suggests that the position 
of this object on the JHK color-color diagram does not arise from an underestimate  
of the reddening correction. This star is associated with complex reflection nebulae
and has a class I-like companion at 6" north that could be contributing
to the observed NIR excesses \citep{fukagawa02,lagage93}.
The early F star located to the left from the B0 reddening line
is BO Cep (HBC735) which has a double peak profile at 
H$\alpha$ line. The 2MASS colors of this object are 
flagged as a source contaminated by nearby star.

We have extracted from the JHK$_s$ colors for stars in our sample from the 2MASS All Sky Survey.
In general, the positions of the Hipparcos
stars and their respective 2MASS sources match within 0.5";
only for the star HIP80473 in US the difference is
larger than 1". The object HIP111104 has JHK$_S$ magnitudes
contaminated by nearby stars.
Figure 3 shows
the JHK color-color diagrams 
of the associations US (a), Per OB2 (b), Lac OB1 (c), Ori OB1a (d) and 
Ori OB1bc (e), and for the probable members of the cluster Trumpler 37 (f) from \citet{contreras02},
indicating the stars with H$\alpha$ in emission.
Stars located in the HAeBe region are
labeled with their respective Hipparcos numbers.
We will discuss individual associations in \S 4.

\subsection{Distances}

Astrometric distances for the associations were obtained 
using the Hipparcos parallaxes 
of the stars in each association. 
We applied a $\chi^2$ test over the distribution 
of distance of the stars in each association, 
changing the theoretical distance until a minimal value of $\chi^2$ 
was obtained by comparing with the observational distance in the sample;
in this way,  we obtain a first guess for the distance of each stellar group. 
We calculated the standard deviation ($\sigma$) 
using the individual distances of the stars and this first guess 
of the distance.
To improve our determination, we rejected 
those values with differences larger than 3 $\sigma$. 
Then we ran the $\chi^2$ test on the improved distribution of parallaxes
to obtain the value of distance reported in column 6 of Table 1.
With our estimate, we reject less than 10\% stars from the
original sample. 
The errors reported in Table 1
are the uncertainty in the mean $\sigma$ ($\frac{\sigma}{\sqrt{N}}$),
where N is the number of stars  used to calculate 
the distances in each association. Except for Lac OB2, our determination 
is in agreement with the previous values of distance calculated using 
Hipparcos data \citep{zeeuw99, brown99b}.

We also estimated photometric distances 
fitting each sample to the ZAMS defined by 
\citet{allen2000} on the B-V, M$_V$
color magnitude diagram. 
We applied a $\chi^2$ test comparing the dereddened values 
of B-V and M$_V$ of the stars in each association with 
the respective values of the ZAMS \citep{allen2000}.
In each case we changed the distance of the association, 
used to calculated M$_V$,
until a minimum value of $\chi^2$ is obtained.  
We used a similar rejection criterion  to the one described 
in the preceding paragraph to improve the $\chi^2$ test. 
In this case, we also rejected 
stars with  B-V$>$0.0 to avoid
stars that have not yet reached the ZAMS.
These estimates are less reliable since the method 
is dependent on the evolutionary status, 
metallicity of the stellar groups, 
presence of multiple systems and photometric calibration 
\citep[e.g. ][]{vandenberg89, pinsonne98, robichon99, carretta99}.
However, except for Ori OB1bc which has a large number of stars 
above the ZAMS because of its youth, the photometric 
and astrometric distances are in good agreement within 
the errors estimated for each stellar group.

\subsection{Hertzsprung-Russell diagrams}

We calculated the stellar luminosity 
using the Hipparcos V magnitude corrected for reddening with
$A_V$ given in the column 5 of Table 2 (\S 3.1),
bolometric corrections from \citetalias{kh95}, and
astrometric distances calculated in \S 3.3.
The effective temperature was determined 
using our spectral types and the calibration 
by \citetalias{kh95}. With these values we plot the 
HRD for each stellar group in Figure 4.
As reference we plot the ZAMS, 
the evolutionary tracks for 0.2, 0.6, 1.0, 1.5, 2.0, 3.0 and 4.0 \msun
from \cite{palla93} and for 5, 9 and 15 {\msun} from \citet{bernasconi96}.  
For each association, we plot the isochrones from \cite{palla93} corresponding
to the age range reported in Table 1.
We derive masses for the stars 
in each stellar group by double interpolation 
on the tracks using the values of \teff 
and luminosities calculated previously. This information 
is shown in Table 2. 
 
\section{Census of Herbig Ae/Be in the associations}

	Table 3 lists the emission line stars 
identified from their spectra in each association. Columns 1, 2 and 3 show the 
Hipparcos name, other name and the association to which the stars belong. 
Columns 4 and 5 give the EW$_\lambda$ of the H$\alpha$ and H$\beta$
lines, respectively (see \S 3.1). Column 6 gives 
the type of star (HAeBe or CBe) based on the 
location of the stars on the dereddened JHK 
color-color diagram (Figure 2, see \S 3.2).
The associated IRAS source, if it exists, is given in column 7. The 
$\beta$ index calculated following the method described in \S 4.1  
is show in the last column. 
This information is used to identify stars with inner disks 
in each stellar group.

\subsection{Upper Scorpius}

Upper Scorpius is the youngest of the three 
subgroups which form the Scorpius-Centaurus 
association, the OB association nearest to the sun.  
The other two subgroups, 
Upper Centaurus Lupus and Lower 
Centaurus Crux with ages of 13 and 10 Myr \citep{brown99a}
extend southward of our telescope pointing limit.
The age of US is about 
5 Myr \citep{degeus89, blaauw91, preibisch99, brown99a, preibisch02}.
The distance reported by \citet{zeeuw99} using Hipparcos 
parallaxes is in good agreement with the astrometric
and photometric distances derived in \S 3.3.
Four objects exhibit emission in the H$\alpha$ line. 
Two of these stars (HIP79476 and HIP81624) are located
in the HAeBe region on the JHK color-color diagram, 
the other two (HIP78207 and HIP80569) are close 
to the CBe region (see Table 3 and Figure 3a).

The SEDs of the H$\alpha$ emission stars 
confirm the classification in HAeBe or CBe based on the
location in the
JHK color-color diagram. \citet{vieira03}
defined an spectral index ($\beta$) using the ratio of the IRAS flux at 12$\mu$m
(F$_{12}$) to the flux in the visual band (F$_V$), 
to discriminate HAeBe from weaker infrared sources as CBe or Vega-like stars. 
Specifically, the index $\beta$ is defined as
\begin{equation} 
    \beta=0.75 \, {\rm log} \, (F_{12}/F_V) - 1
\end{equation}
HAeBe have spectral index {$\beta\gtrsim-2.0$}  \citep{vieira03}.

We calculated fluxes for each H$\alpha$ emission star 
using the B, V, I$_c$ \citep{esa97} and J, H, K$_S$ \citep{cutri03} 
magnitudes corrected by reddening and the 
flux density for Vega \citep{allen2000} at each band. Figure 5
shows the SEDs of the emission line star in US.
Stars labeled as HAeBe from the JHK color-color diagram, 
have {$\beta>-2$},  characteristic of star with disks \citep{vieira03}. 

All the stars in the sample, including the HAeBe, 
are located near or on the main sequence  
in the HRD (Figure 4a), indicating that the members 
of this sample have common properties. Only the star HIP80569,
labeled as a CBe, has a slightly different position 
in this diagram, and its membership may have to be reviewed.
Most of the lower mass stars in the sample follow the isochrones
between 4 and 6 Myr, in agreement with the age reported 
for the association.

We identify the Herbig Ae/Be stars with stars with
inner optically thick disks (\S 1), and use their
numbers relative to the total to find the
inner disk frequency (\%F$_{disk}$) in each association.
To analyze the fraction of stars with inner disks, 
we limited the sample to the 
spectral type range B5-F0. The presence of disks around stars 
earlier than B5 is not well established, possibly 
due to the rapid evolution of these objects
which disperse the surrounding dust and gas 
in about 1 Myr \citep{fuente02, natta00}. 
Among the 93 objects observed in US, 13 have spectral types earlier than B5,
20 have spectral types later than F0, and 60 have spectral types
between B5 and F0  (N$_{B5-F0}$); 
of these, only two objects can be 
labeled as stars with inner disks (N$_{disk}$). 
This indicates that in US 3.3 {$\pm$} 1.3\% of the intermediate 
mass stars have inner disks. The error is calculated from 
the ratio  \%F$_{disk}$/$\sqrt{N_{disk}}$.

We explored the NIR properties of the 11 stars
classified as early members 
(5 stars have spectral type between B5 and F0)
by \citet{zeeuw99} for which we did not obtain spectra.
On the JHK color-color diagram, 
these stars fall to the left of the reddening vector from the B0 star,
near the dwarf standard sequence. Since these data are not corrected by 
reddening, 
they must have small values of A$_V$ and no NIR excess. Thus, these 
stars are unlikely to have inner disks.
With these stars, \%F$_{disk}$ = 3.1 , which is within
the errors of our original estimate. 

\subsection{Perseus OB2}
 
Using photometric studies, \citet{gimenez94} found 
an age 10-15 Myr for Per OB2; \citet{brown99a} give an age 
between 4 and 8 Myr citing ages reported by different 
authors \citep[see also, ][]{zeeuw99}. 
However, the ages reported for the open cluster IC348, 
a concentration of stars embedded 
in Per OB2, are in general younger: 5-20 Myr
\citep{strom74}, 5-7 Myr \citep{lada95}, 
3-7 Myr \citep{trullols97}, and 1.3 - 3.0 Myr 
\citep{herbig98}. Some authors suggest that 
the Per OB2 association consists of two subgroups 
\citep{herbig98, hakobyan00, belikov02a, belikov02b}.
In particular, Herbig (1998) uses spectroscopic 
considerations based on the presence of H$\alpha$ to
suggest that IC 348 is projected upon an older 
population of stars. 
This conclusion was  supported by \citet{belikov02a} 
using astrometric data. Since most of the stars in 
our sample are located outside of IC348, we adopt 
the age given by \citet{brown99a}, 
which is older that the age estimated by \citet{herbig98}
for IC348.  The distance reported 
by \citet{zeeuw99} is in good agreement with 
the astrometric and photometric distances derived in \S 3.3.
 
In our sample of 40 stars, which include all 
the early members listed by \citet{zeeuw99},
we do not find stars with emission lines or 
NIR excess (see Figure 3b). Thus, 
the inner disk frequency in this stellar group 
is 0$\pm$4.3, where the error was calculated
assuming one HAeBe in our sample. 
However, the lack of stars
with inner disks in Per OB2 could 
be due to the low number of objects in the sample 
(N$_{B5-F0}$=23). 

Figure 4b shows the HRD for Per OB2.
The five stars later 
than G0 are unlikely to be members of Per OB2.
We also show
the isochrones for 4 and 8 Myr, but 
the lack of low mass stars prevents us 
from making a firm statement regarding 
the age of the association.

\subsection{Lacerta OB1}

Lacerta OB1 is a moderately sparse group. 
Using proper motions and radial velocities,
\citet{blaauw58} divided the region into 
two subgroups, an older and more sparse 
group extended to the northeast and named 1a,
and other more concentrated group in the 
vicinity of the star HIP111841 (10 Lac). 
The photometric ages derived for 1a and 
1b are 16 and 12 Myr, respectively \citep{blaauw64,brown99a}.  
Using Hipparcos parallaxes, \citet{zeeuw99} 
calculated a distance which is significantly 
smaller than previous estimates \citep{blaauw64,lesh69,crawford76}.
The astrometric distance calculated in \S 3.3 for 
this association is about 3$\sigma$ 
larger than the value given by \citet{zeeuw99}, and 
in better agreement with our photometric 
estimate and the values given by others authors.

Out of the 82 stars observed in this work, 27 have spectral types 
earlier than B5, 3 have spectral types later than F0 and 
52 have spectral types between B5 and F0; 
only four of these objects exhibit emission at H$\alpha$.
All these objects are located in 
the CBe region on the JHK color-color
diagram (Figure 3c), so we conclude that do not detect
any star with an inner disk. 

The H$\alpha$ emission 
objects do not have associated IRAS sources. However, as 
the IRAS survey is complete to about 0.4, 
0.5, 0.6, and 1.0 Jy at 12, 25, 60, 
and 100 $\mu$m \citep{iras88}, we can calculate 
an upper limit for the  $\beta$ index; 
for the emission line objects the index is always 
less than -2.0, confirming the results derived
from NIR colors. In Figure 6, we plot 
the SEDs for HIP110476, HIP112148, 
HIP111546  and HIP113226 with the 
upper limit of the IRAS fluxes.    
We then derive an inner disk frequency of 0 $\pm$ 1.9 \% in Lac OB1. 
The statistical error was calculated  assuming 
the presence of one HAeBe star.

Only one star listed by \citet{zeeuw99} as an early type member 
was not observed in this work (HIP111841). 
However, the colors of HIP111841, not corrected for reddening,
fall near the standard main sequence 
in the JHK diagram.
This indicates 
that this star does not have  NIR colors characteristic of 
HAeBe and does not change the results obtained previously.

\subsection{Orion OB1a}

Orion OB1a is the older group of the Orion OB1 association (see
\S2.2), one of the largest 
and nearest regions with active star formation. 
The age reported for this group 
ranges from 4 Myr \citep{lesh68} to 12 Myr
\citep{blaauw91,brown94}. Using photometric 
variability and spectroscopic confirmation 
to select low mass members belonging to OB1a, \citet{briceno04}
calculated an age of 7-10 Myr for this stellar group.
We adopt this age for Ori OB1a. 
The distance reported by \citet{brown99b} using the average parallax of 61 
Hipparcos stars is in good agreement with our
estimates (see \S 3.3).

In our sample of 114 stars, we have 80 stars with 
spectral types in the range B5 and F0, 25 stars with spectral types
earlier than B5 and 9 objects with spectral types later than F0.  
Six stars in our sample show emission at H$\alpha$ line 
(HIP25258, HIP25299, HIP25302, HIP25655, HIP26476
and HIP26481). 
In addition, in the star HIP25299
we find that H$\alpha$ is in absorption but filled in when
compared to an A8 standard.
An H$\alpha$ emission line with relative
small EW$_\lambda$ is visible in the higher resolution 
spectra of this star obtained by the EXPORT consortium \citep{merin04}.
In addition, the EXPORT multi-epoch spectra show variability 
in this line. So, we have included  HIP25299 as
an emission line star.
The emission line stars show similar
locations on the HRD (Figure 4d)
as other members of
the association.

Figure 3d shows that HIP25299 and HIP25258
are located in the HAeBe region on the JHK color-color 
diagram, while the other four emission 
line stars are likely to be CBe.
In Figure 7 we plot the SEDs for the emission line 
stars. The $\beta$ index calculated for these stars 
confirms the results obtained previously; 
namely, stars located in the CBe region 
(HIP25302, HIP25655, HIP26476 and HIP26481)
on  the JHK color-color diagram have $\beta$ 
characteristic of stars without inner disks 
($<$ -2.0), while the $\beta$ index of HIP25299 and HIP25258
confirms the presence of inner disks.
From this analysis we obtain an inner disk frequency of 
2.5 $\pm$ 1.8 \% for Ori OB1a.   

\subsection{Orion OB1bc}

As discussed in \S 2.2,
we have combined the associations Ori OB1b and Ori OB1c
as defined by \citet{warren77a} into one. These authors reported ages of 
5.1 Myr and 3.7 Myr for Ori OB1b and Ori OB1c, 
respectively. Later, \citet{blaauw91} 
estimated ages of 7 Myr for Ori OB1b and 3 Myr for Ori OB1c. 
In contrast, , \citet{brown94} suggest 
that Ori OB1b is younger than Ori OB1c, deriving ages 
of 1.7$\pm$1.1 and 4.6$\pm$2.0, respectively. 
However, OB1c is more spatially related to 
the youngest sub-association Orion OB1d 
\citep[~1Myr; ][]{hillenbrand97}  and contains large quantities of
dust and gas (see Figure 1). Therefore, Ori OB1c could be co-eval 
or younger than Ori OB1b, 
which has an age of 3-5 Myr \citep{briceno04}.
In any event, in this contribution we assume a range of 
age that includes most of the age estimates for both Ori OB1b 
and Ori OB1c, 3.5$\pm$3 Myr. Using Hipparcos data, 
\citet{brown99b} found that Ori OB1b and Ori OB1c are 
located at similar distances, 440 pc for 1b and 460 
pc for 1c. These values are in good agreement with our 
astrometric distance calculated for the combined 
group Ori OB1bc (\S 3.3).

In our sample of 110 stars, we have 77 stars with 
spectral types in the range B5 and F0, 26 stars with types
earlier than B5, and 7 objects with types later than F0. 
We detect six stars with H$\alpha$ in emission.
Figure 3e shows the position of the emission line
stars on JHK color-color diagram. 
Objects HIP27452 and HIP27842 clearly appear in the CBe
region on this diagram, while stars HIP25258, HIP25299
and HIP25302 have NIR excesses characteristic of HAeBe. 
HIP26500 is located between
the CBe region and the HAeBe region.
This star is a double system composed by stars with 
similar spectral types \citep{guetter76}. 
An IRAS source is located 13" form HIP26500, and there 
are other possible objects associated with this IRAS source.
We derived the SED and $\beta$ index
assuming that this IRAS source is associated with HIP26500 
($\beta$=-2.2) and using the completeness limit of the IRAS catalog 
($\beta <$-2.5).  However, the IRAS fluxes do not help to determine
if this star is a CBe or a HAeBe star, because the $\beta$
index is close to the limit defined by \citet{vieira03}.
For the other stars located in 
the CBe star region on the JHK color-color diagram,
HIP27452 and HIP27842,
the index $\beta$ confirms the absence of disks.
The remaining H$\alpha$ emission stars, HIP26752,
HIP27059 and HIP26955, show clear evidence of being HAeBe. 
Figure 8 shows the SED for the emission line stars 
in Ori OB1bc. Using the bona fide HAeBe, the fraction of 
inner disks present in Ori OB1bc is 3.8 $\pm$ 2.2 \%.
If we include the object HIP26500 as HAeBe, 
the fraction increases to 5.1 $\pm$ 2.6 \%. 

A small number of stars located around l=206$^\circ$ and
b=-24$^\circ$ are not spatially related to dust or 
gas (Figure 1). It is possible that this group has different
age and distance  than those of Ori OB1bc; 
however, since there few of these stars, they are not likely to affect the inner disk frequency 
determination.

Figure 4e shows that most of the stars are located between the isochrones corresponding
to the age range estimated for this stellar group ($3.5 \pm 3$ Myr);
the emission line stars show similar locations on 
the HRD as other members of the association. 

\subsection{Trumpler 37}

	We studied the inner disk frequency in Trumpler 37 (Tr 37) 
using the spectral types, extinctions and membership defined by 
\citet{contreras02}. Tr 37 lies in the Cep OB2 association at 
a distance of 900 and with an age of 3-5 Myr.
The spectroscopy was done using the same instrumental setup 
as for the other stellar groups studied in
this contribution (see \S2.2).

Out of 66 stars defined by \citet{contreras02} as probable members, 
56 have spectral types between B5 and F0. Only 3 of these stars 
exhibit emission at H$\alpha$, MVA 437, MVA 426 and KUN 314S. 
The star KUN 341S is presumably heavily veiled by accretion 
onto the central star , and no spectral type could be assigned 
\citep{contreras02}. The other two stars were classified as B7.

We used 2MASS data to infer the presence of inner disks 
in the H$\alpha$ emission objects. We corrected for reddening the 
JHK$_S$ magnitudes using the values of A$_V$ calculated by \citet{contreras02}.
The star MVA 437 falls in the CBe region while the star
MVA 426 is located in the HAeBe region on the JHK color-color diagram
(Figure 3f). The star KUN 314S does not have a determination of A$_V$ in \citet{contreras02} 
due to  its high veiling. However, we can use 
the average extinction A$_V$=1.67$\pm$0.42
for Trumpler 37 calculated by
\citet{aurora04} 
from a sample of low mass stars in the cluster
to correct for reddening the J-H and H-K$_S$ colors.
To support the use of the mean extinction for this star,
we used this value of A$_V$ to correct the
near infrared colors of the northern companion of the
system, KUN 314N, and found the colors to correspond
to those of an early G type star
on the standard main sequence. This type is consistent
with the spectral type G2 that we obtained from
analysis of an unpublished spectrum of this star,
which does not exhibit any emission features.
With this reddening correction, the colors of KUN 314S
are located between the CBe 
region and the HAeBe region, so we do not have 
clear evidence 
for an inner disk around this star. Also, this star 
does not have an associated IRAS source so we can 
only estimate an upper limit for $\beta$ (see Table 3 and
Figure 9). If we assume this object to be a HAeBe star, 
the resulting inner disk frequency for intermediate mass
stars in Tr 37 is 
$4.3 \pm 3.0$\%. Otherwise we obtain a value of $2.2 \pm 2.2$\%

\section{The inner disk frequency in intermediate mass stars}

In sections 4.1 to 4.6 we determined the relative
numbers of Herbig Ae/Be stars in each OB association,
and identifying these stars with stars with inner
disks, we estimated
the inner disk frequency in intermediate
mass stars. In Figure 10, we plot
this frequency as a function of the age of the stellar group. 

We compare our determinations of inner disk
frequencies with similar quantities in low mass stars.
In Figure 10 we show estimates for NGC2024, the Trapezium, IC348, NGC2264
and NGC2362 from \citet{haisch01}, for Taurus \citepalias{kh95},
for Chameleon I \citep{gomez01}, and for NGC7129 \citep{gutermuth04}.

We also include in Figure 10 the disk frequencies
in low mass stars in Ori OB1a and 1b,
which we calculate from the number
of stars in \citet{briceno04} which show excesses relative to the
main sequence in the JHK diagram.
We assume that
a star has an excess if
$\rm (H-K)_0 - (H-K)_{KH}$ $>$ $\rm 2\sigma_{KH}$, 
where $\rm (H-K)_0$ is the dereddened color, $\rm (H-K)_{KH}$ is the intrinsic color 
for the spectral type of the star taken from
\citetalias{kh95}, and
$\rm \sigma_{KH} = 0.04$, which is the typical dispersion
for Weak T Tauri stars (or disk-less stars, Kenyon \& Hartmann 1995).
Using this criterion, we find a disk frequency of $7.4 \pm 3.4$\%
for Ori OB1a and $17.4\pm 3.6$\% for Ori OB1b.

Parenthetically, we note that the
frequencies determined from near infrared
excesses in Ori OB1 are lower than the frequencies
estimated from the relative numbers of CTTS in these
associations, 10 $\pm$ 4\% for Ori OB1a and 23 $\pm$ 4\%, by
\citet{briceno04}. These latter fractions refer to
disks which are still accreting mass onto the star, as
indicated by the presence of strong H$\alpha$ emission
characteristic of the CTTS
(Muzerolle, Calvet, \& Hartmann 2001).
The lower fraction of inner dusty disks compared to accreting disks
is consistent with our findings  in \citep{calvet04b}, where we show
evidence for significant dust evolution in the disks of
Ori OB1, and suggests that regions
like Ori OB1 harbor a number of transition objects like TW Hya
\citep{calvet01,uchida04},  with 
significant clearing of small dust particles in their inner disks.

The inner disk frequency in the Herbig Ae/Be stars
is much lower than in the low mass stars
in the age range 3 - 15 Myr.
In addition, we do not see any indication of the strong decrease in
disk frequency at ages 3-5 Myr in the low mass stars; rather, the inner
disk fraction in HAeBe starts out already very low at 3 Myr ($\sim
2-5$\%) and then slowly decreases with time up to $\sim 15$ Myr when it
has fallen to essentially zero.
The lifetimes of disks seem to depend on the stellar mass, as
suggested by \citet{haisch01}; for intermediate mass stars, 
the inner disks dissipate more quickly than for the 
low mass stars. 

Several possibilities may explain the mass
dependence of the disks lifetimes. 
\citet{calvet04a} extended the correlation between mass accretion 
rate (\.{M}) and stellar mass (M$_*$) from \citet{muzerolle03a}
to the intermediate mass stars, up to $\sim$ 3.5\msun;
in particular, they found that \.{M} $ \propto M_*^2$.
On the other hand, the ratio of 
disk mass (M$_D$) to M$_*$ is found to be approximately
constant for stars with spectral type
A0-M7 \citep{natta00,fuente03}. If the lifetimes of disks are 
determined by viscous evolution, which
controls the accretion rate onto the star, the time 
for the disk to disappear
by accretion processes can be estimated as the ratio M$_D$/\.{M}, so
the viscous lifetime of the disk is $\propto M_*^{-1}$. 

Grain growth and settling toward the midplane in the disk can decrease
the height  of the inner wall of the disk and so decrease the NIR contribution
from the inner disk \citep{calvet04a}.
Following \cite{dullemond04b}, 
in absence of turbulent stirring, the settling speed
is given by
\begin{equation}
	V_{sett} \: \propto \: \frac{M_*\: z \: m}{r^3 \: \rho(z)}       
\end{equation}
where $z$ and $r$ give the vertical and radial location of 
a particle with mass m that is settling, and $\rho(z)$ is the gas density
in the disk, 
which decreases exponentially with z away from the midplane. 
$V_{sett}$ is proportional to the mass of the star, 
so settling is expected to be more effective
in the inner disks of HAeBe than 
in their low mass counterparts. 
Thus, the processes which lead to the dissipation
of inner disk, that is, viscous and dust evolution are expected
to occur in shorter time-scales in more massive stars.

\section{Summary and Conclusions}

	In this contribution we studied the presence of inner disks 
around intermediate mass pre-main sequence stars in 5 stellar groups 
with range of ages from 3 to 16 Myr. The stars were selected from 
the Hipparcos catalog. The astrometric membership for US, Per OB2 
and Lac OB1 associations was determined  from \citet{zeeuw99}. 
The stars in Orion OB1 were selected as members using 
astrometric and photometric constrains \citep{brown94,brown99b} 
and divided in two groups, 
Ori OB1a and  Ori OB1bc, with different ages and distances. 
We obtained spectra for 440 stars in these
OB associations, from which
we have determined 
spectral types, visual extinctions, \teff, luminosities and the presence of 
emission at H$\alpha$. We also included data for the cluster Trumpler 37 in
the association Cep OB2, with data
from \citet{contreras02}.

	We compared the dereddened J-H and H-K colors of a samples of HAeBe 
stars \citep{hernandez04} and CBe \citep{yudin01} 
in order to estimate the loci of these objects, showing that the HAeBe 
and the CBe occupy well defined different regions on this 
diagram. 

We identified the Herbig Ae/Be stars in each association
using the following criteria (1) H$\alpha$ emission,
(2) location in the HAeBe region of the JHK diagram,
(3) strong IRAS fluxes with \citep[$\beta$ index $>$ -2; ][]{vieira03}.
We identified 2 HAeBe in Upper Scorpius,
0 HAeBe in Perseus OB2, 0 HAeBe in Lacerta OB1, 
2 HAeBe in Orion OB1a, 4 HAeBe in Orion OB1bc
and 2 HAeBe in Trumpler 37.
From these identifications, we estimate
a disk frequency of 3.3$\pm$1.7\%, 0$\pm$4.4\%, 0$\pm$1.9\%, 
2.5$\pm$1.7\%, 5.1$\pm$2.0\%, 4.3$\pm$1.8\%,  respectively.
We have compared the inner disk frequency in intermediate
mass stars (HAeBe star) with that 
for low mass stars in a number of
stellar groups taken from the literature, and
estimated by us for Ori OB1 from the sample of \citet{briceno04}.
We find that stars with inner disks 
are more frequent in low mass stars than in intermediate mass stars.
Particularly, for the youngest associations (Ori OB1bc and Tr 37)
the inner disk frequency for intermediate mass stars is $\sim$ ten times 
lower than the inner disk frequency reported for low mass stars.
Although the inner disk frequency in intermediate
mass stars tends to decrease slowly with age,
we find no evidence of the  
strong decrease in the inner disk frequency 
observed in low mass stars at 3-5 Myr. 
Studies of intermediate mass stars
in stellar groups younger than 3 Myr
are required to determine in which age range, if any,
a similar strong decrease with age occurs. 
We suggest that the lower inner disk frequency
in intermediate mass stars is a result of more
rapid mechanisms of inner disk dispersal (accretion,
dust growth and settling toward the midplane).

This contribution suggests that most of the stars in the 
range of mass of the HAeBe disperse their inner disks in 
a timescale shorter than 
3 Myr; in contrast, the inner disk frequency of low mass stars in this
age range is 
larger than 50\%. Samples of younger stellar
groups are required to determined when the most of the
intermediate mass stars disperse their inner disks.

\section{Acknowledgments}

	We thank Francesco Palla to provide us the isochrones and 
evolutionary tracks of PMS stars, Bruno Merin for sending us the multi-epoch
spectra of the star HIP26955, Thomas Megeath for  comments about 
the association NGC 7129, Robert Wilson for providing with the
$^{13}$ CO map of the Orion region, and Charlie Lada for useful
comments. We  also thank Susan Tokarz 
of the SAO Telescope Data 
Center for carrying out the data reduction, and Michael Calkins
for obtaining some of the spectra. 
This publication makes use of data products from the Two Micron All
Sky Survey,  which is a joint project of the University of
Massachusetts and  the Infrared Processing and Analysis
Center/California Institute of Technology, funded by the National
Aeronautics and Space Administration and the National Science Foundation.
This work was supported in part by NASA grants NAG5-9670
and NAG10545, NSF grant AST-9987367 and 
grant No. S1-2001001144 of FONACIT, Venezuela.

\begin{deluxetable}{ccccccc}
\tabletypesize{\scriptsize}
\tablewidth{0pt}
\tablecaption{Nearby Associations\label{tab:Asso}}
\tablehead{
\colhead{Name} & \colhead{D$_{Ref}$} & \colhead{Age} & \colhead{N} & \colhead{N$_{spt}$} & \colhead{$D_{\pi}$} & \colhead{$D_{CMD}$} \\
\colhead{ } & \colhead{ pc } & \colhead{Myr} & \colhead{ } & \colhead{ } & \colhead{pc} & \colhead{pc} 
}
\startdata
Upper Scorpius & 145 $\pm$ 2  &   5   & 120 & 93 & 144 $\pm$ 3  & 146 $\pm$ 5\\ 
Per OB2       & 318 $\pm$ 27 & 4$-$8 & 41  & 40 & 313 $\pm$ 13 & 300 $\pm$ 15\\ 
Lac OB1       & 368 $\pm$ 17 & 16    & 96  & 82 & 418 $\pm$ 15 & 438 $\pm$ 20\\ 
Ori OB1a      & 336 $\pm$ 16 & 7$-$10    & 124 & 114 &  335 $\pm$ 13 & 315 $\pm$ 8 \\ 
Ori OB1bc*    & 438-462 & 1$-$7   & 121 & 111 &  443 $\pm$ 16 & 392 $\pm$ 20 \\ 
\enddata 
\tablecomments{~}
\tablenotetext{*}{The association Orion OB1bc have 2 sub-groups 
with similar distances and ages ( Ori OB1b and Ori OB1c see \S 2.2)}
\end{deluxetable}

\begin{deluxetable}{cccccccccccc}
\tabletypesize{\scriptsize}
\tablewidth{0pt}
\tablecaption{Stellar Properties \label{properties}}
\tablehead{
\colhead{Hipparcos} & \colhead{Name} & \colhead{SpT} & \colhead{Error} & \colhead{A$_V$} &
\colhead{$\sigma$(A$_V$)} & \colhead{J} & \colhead{J-H} &
\colhead{H-K$\rm{_S}$} & \colhead{log(\teff)} &
\colhead{log(L/$\lsun$)} &\colhead{M/\msun}\\
%\colhead{Hipparcos} & \colhead{Name} & \colhead{SpT} & \colhead{Error} & \colhead{A$_V$} &
%\colhead{$\sigma$(A$_V$)} & \colhead{J} & \colhead{J-H} &
%\colhead{H-K$\rm{_S}$} & \colhead{log(\teff)} &
%\colhead{log(L/$\lsun$)} &\colhead{M/\msun}
}
\startdata
\cutinhead{Upper Scorpius}
HIP76071 & HD138343 & B9 & 1 &   0.28 &   0.14 &   7.11 &   0.03 &   0.08 &   4.02 &   1.64  &	2.5 \\
HIP76310 & HD138813 & A1 & 2 &   0.09 &   0.23 &   7.16 &  -0.02 &   0.01 &   3.96 &   1.43  &	2.2 \\
HIP78207$^1$ & HD142983 & B1 & 2 &   0.62 &   0.41 &   5.10 &   0.27 &   0.24 &   4.44 &   3.66  &	12.8 \\
HIP79439 & HD145631 & B9 & 1 &   0.63 &   0.15 &   7.05 &   0.00 &   0.11 &   4.02 &   1.65  &	2.5 \\
\cutinhead{Perseus OB2} 
HIP14145 & BD+42 684 & F8 & 1 &   0.23 &   0.13 &   8.60 &   0.21 &   0.06 &   3.79 &   1.19  &	     2.1 \\
HIP17172 & HD22765  & A5 & 2 &   0.31 &   0.24 &   8.83 &   0.01 &   0.10 &   3.91 &   1.37  &	     2.0 \\
HIP17313 & HD22951  & B1 & 1 &   0.65 &   0.18 &   5.00 &  -0.07 &   0.01 &   4.41 &   4.26  &	    13.4 \\
\enddata 
\tablecomments{Table \ref{properties} is published in its entirety in the electronic edition of the {\it Astrophysical Journal}. A portion is shown here for guidance
regarding its form and content.}
\tablenotetext{1}{ Stars with emission in H$\alpha$ line (see table 3)}

\end{deluxetable}

\begin{deluxetable}{ccccccccc}
\tabletypesize{\scriptsize}
\tablewidth{0pt}
\tablecaption{Emission line stars \label{tab:emi}}
\tablehead{
\colhead{Hipparcos} & \colhead{Name} & \colhead{Assoc} & \colhead{EW$_\lambda$[H$\alpha$]} & \colhead{EW$_\lambda$[H$\beta$]} & \colhead{JHK} & \colhead{IRAS} & \colhead{index} & \colhead{Inner} \\
\colhead{ } & \colhead{ } & \colhead{(\AA)} & \colhead{(\AA)} & \colhead{ loci } & \colhead{} & \colhead{$\beta$} & \colhead{Disk} 
}
\startdata
HIP78207 & 48 Lib  & US      & -21.0 & -0.1    & Be    & 15553-1408 & -3.2 & N \\ 
HIP79476 & V718 Sco  &  US      & -0.6  & \nodata & HAeBe & 16102-2221 & -1.7 & Y \\ 
HIP80569 & HD148184  & US      & -36.4 & -3.9    & Be    & 16241-1820 & -3.0 & N \\ 
HIP81624 & V2307 Oph  & US      & -11.9 & \nodata & HAeBe & 16372-2347 & -1.6 & Y\\
HIP110476 & BD+42 4370 & Lac OB1 & -18.7 & -0.5   & Be    &  \nodata   & $<$-2.4 & N \\ 
HIP111546 & HD214167  & Lac OB1 & -12.9 & -0.1   & Be    &  \nodata   & $<$-3.3 & N \\ 
HIP112148 & HD215227  & Lac OB1 & -23.1 & -1.2   & Be    &  \nodata   & $<$-2.7 & N \\ 
HIP113226 & HD216851 & Lac OB1 & -41.2 & -2.4   & Be    &  \nodata   & $<$-2.8 & N \\
HIP25258 & HD287823 & Ori OB1a & -0.4  & \nodata & HAeBe & 05215+0225 & -2.0 & Y \\ 
HIP25299 & V346 Ori & Ori OB1a & -0.1  & \nodata & HAeBe & 05221+0141 & -2.0 & Y \\ 
HIP25302 & V1086 Ori & Ori OB1a & -3.8  & \nodata & Be    & 05221+0148 & -3.1 & N \\ 
HIP25655 & V1372 Ori & Ori OB1a & -26.0 & -1.4    & Be    & \nodata    & $<$-2.9 & N \\
HIP26476 & HD37330 & Ori OB1a & -5.4 & \nodata  & Be    & \nodata    & $<$-2.8 & N \\ 
HIP26481 & HD37342 & Ori OB1a & -5.6 & \nodata  & Be    & \nodata    & $<$-2.6 & N \\
HIP26500 $^2$& HD37371 & Ori OB1bc & -4.3  & \nodata & \nodata & \nodata & $<$-2.5 & ? \\ 
HIP26752 & HD37806 & Ori OB1bc & -24.5 & \nodata & HAeBe & 05385+0244 & -2.0 & Y \\ 
HIP26955 & HD38120 & Ori OB1bc & -21.6  & -2.1   & HAeBe & 05407+0501 & -3.1 & Y \\ 
HIP27059 & V351 Ori & Ori OB1bc & -0.9  & \nodata & HAeBe & 05417+0007 & -2.9 & Y \\
HIP27452 & HD38856 & Ori OB1bc & -13.9 & -0.6    & Be    & \nodata    & $<$-2.8 & N \\ 
HIP27842 & HD39557 & Ori OB1bc & -24.0 & -0.7    & Be    & \nodata    & $<$-2.7 & N \\
\nodata & MVA 426 & Tr 37      & -6.9  & \nodata & HAeBe & 21365+5713 & -2.0  & Y \\
\nodata & MVA 437 & Tr 37      & -8.3 &  \nodata & Be    & \nodata    & $<$-2.1 & N \\ 
\nodata $^2$ & KUN 314S & Tr 37     & -38.1 & -3.9    & \nodata    & \nodata    & $<$-1.8 & ? \\
\enddata 
\tablenotetext{~}{ The stars in Per OB2 do not show H$\alpha$ in emission}
\tablenotetext{2}{ JHK loci between the HAeBe region and the CBe region}
\end{deluxetable}

\begin{figure}
\includegraphics[scale=0.6,angle=-90]{Hernandez.fig1.ps}
\caption{Spatial distribution of the Hipparcos stars in the Orion OB1 association. 
A map of $^{13}$CO \citep{bally87} is shown in gray scale halftone, covering a range 
of integrated $^{13}$CO emissivity from 0 to 35 K km$s^1$. The isocontours are estimators of galactic 
extinction (for A$_V$=1,2,3 and 4 mag) from the map of Dust Infrared Emission \citep{schlegel98}.
The dashed lines are the limits of the sub-association from \citet{warren77a}  for 
1a (open triangles), 1b (open circles inside the dashed box) and 1c (open circles outside of the dashed box). 
A subset of stars in the region defined for Ori OB1a \citep{warren77a}  
(open triangles surrounded by open circles), is spatially associated 
with molecular gas and dust; this subset is more likely related to the younger 
subassociation Ori OB1c.}
\label{fig:OriOB1a}
\end{figure}

\begin{figure}
\epsscale{0.7}
\plotone{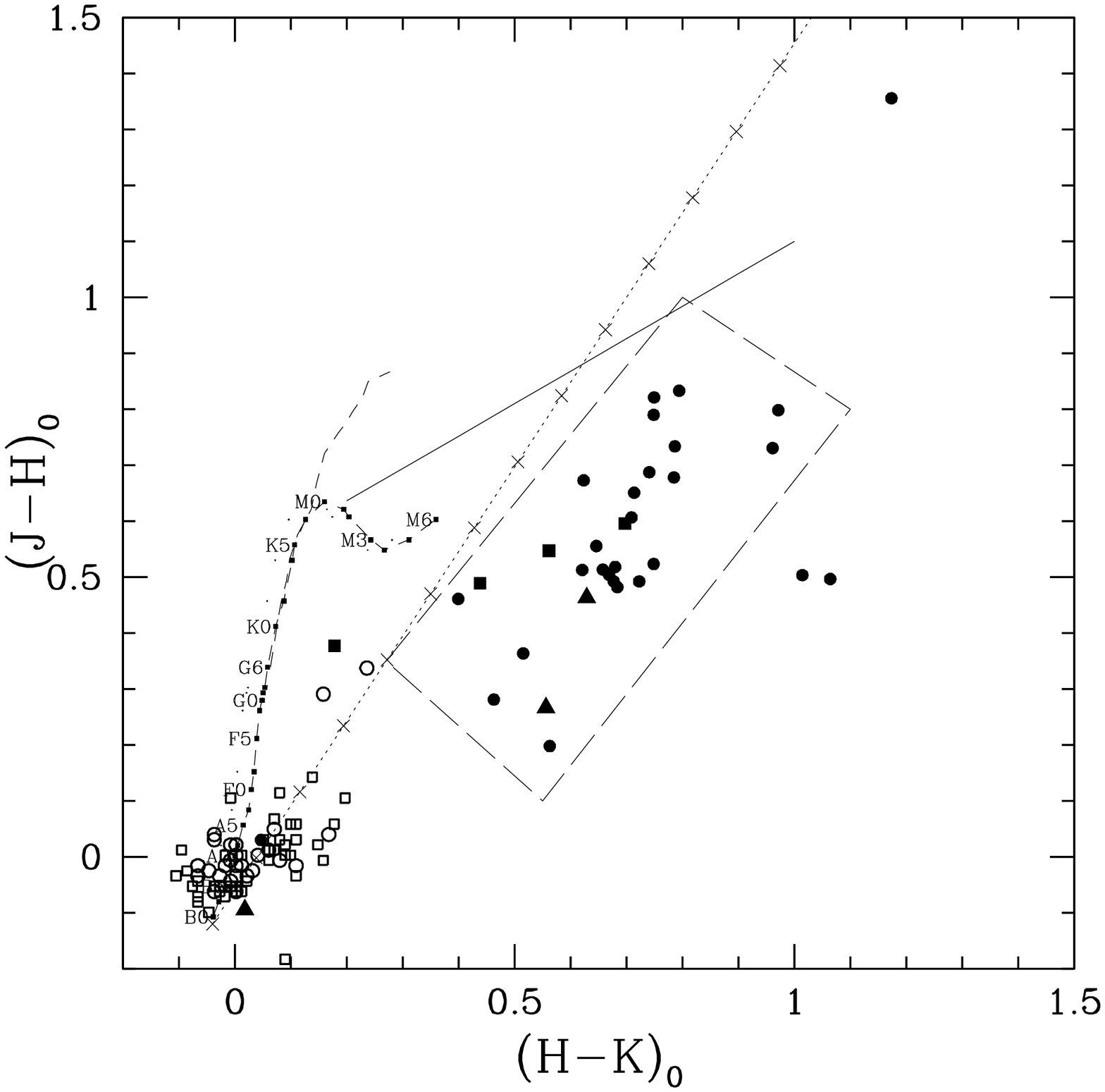}
\caption{Loci of the HAeBe and the CBe on the JHK color-color diagram. 
The standard sequences from \citet{bessell88} are shown in dashed lines.
Colors of the HAeBe sample, taken from \citet{hernandez04}, corrected for reddening, fall
to the right of the reddening line of a standard B0 star 
(dotted line). This line is calculated for R$_V$=3.1 with
the reddening law of \citet{ccm89}; the tick-marks show
intervals of A$_V$= 1. The reddening line
for R$_V$=5.0 has similar slope \citep{ccm89}, but larger spacing between
tick-marks.
The sample of HAeBe is separated in three spectral type ranges: 
earlier than B5 (solid filled triangles), between B5 and F0 (solid filled circles) and later than F0 (solid filled squares).  
The colors of the sample of CBe from \citet{yudin01},
corrected for reddening, concentrate in a  region near of the
 blue end of the standard main sequence.
 The CBe are also separated in spectral type:
B5 or later (open circles), earlier than B5 (open squares). 
As a reference,
we plot the loci of CTTS (solid line) defined by \citet{meyer97}. It is apparent 
that the HAeBe and the CBe occupy different regions on the JHK diagram.
The HAeBe are concentrated in a region (long dashed lines) on the JHK color-color diagram
approximately defined by the vertices
(J-H,H-K)=(0.27,0.35); (0.80,1.00); (1.10,0.80); (0.55 0.80). 
All the measurements were transformed to the CIT system (see \S3.2)}.  
\label{fig:Locus}
\end{figure}

\begin{figure}
\epsscale{1.0}
\plotone{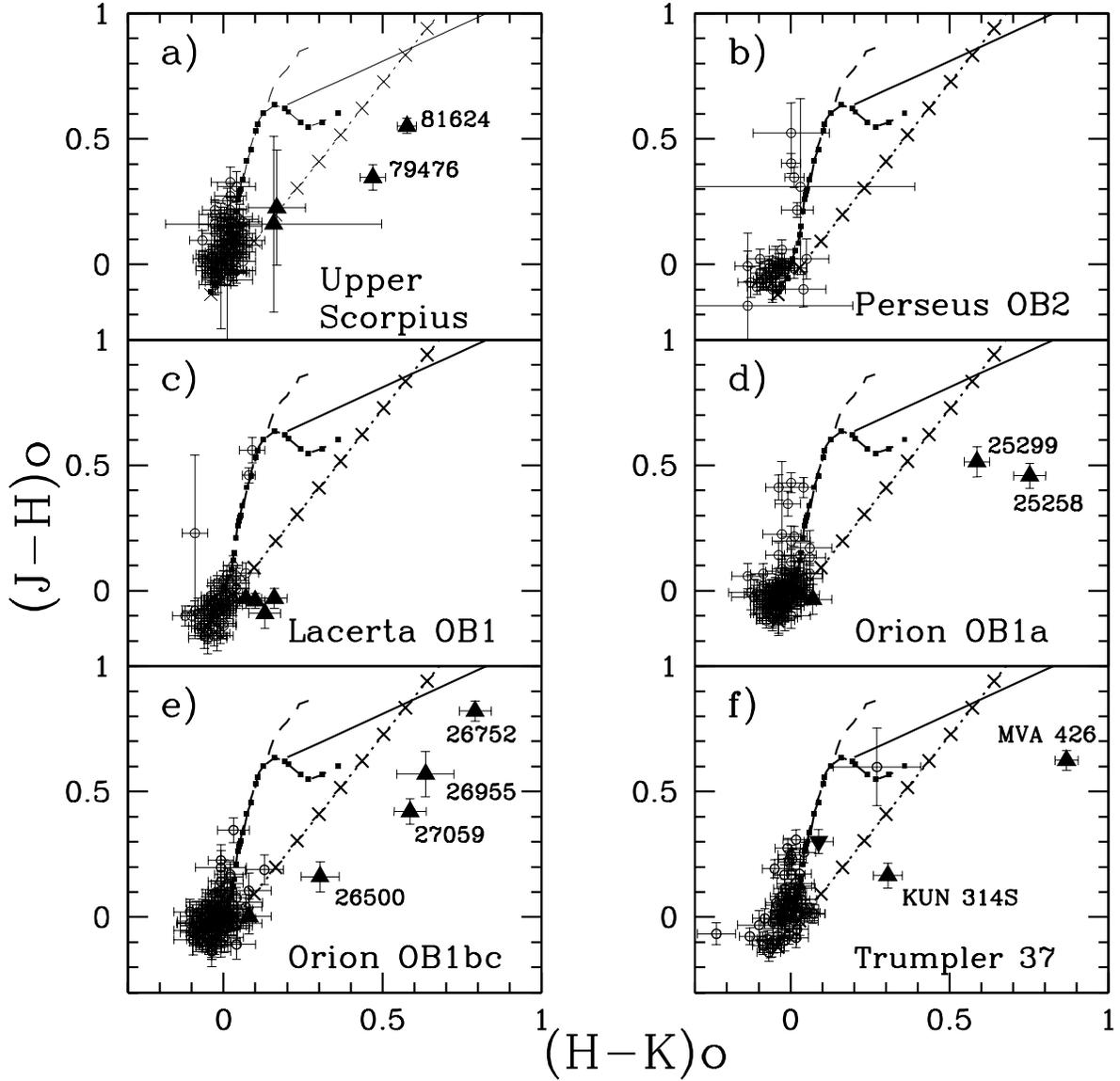}
\caption{ Dereddened J-H and H-K colors of the
early type stars in the OB associations studied in this work (\S 4).
Stars with emission in H$\alpha$ are plotted as solid triangles. 
Stars without emission lines (open circles) concentrate on the
main sequence or in the CBe region (see Figure 2).
The inverse solid triangle in panel f is the star KUN314N (\S 4.6)
Other symbols as in Fig. 2.
Objects located in the HAeBe region are labeled with their
number in the Hipparcos catalog. The error bars are estimated by propagating both the
error in our spectral type determination and the error from the 2MASS
catalog.  
}
\label{fig:JHKD}
\end{figure}

\clearpage
\begin{figure}
\epsscale{1.0}
\plotone{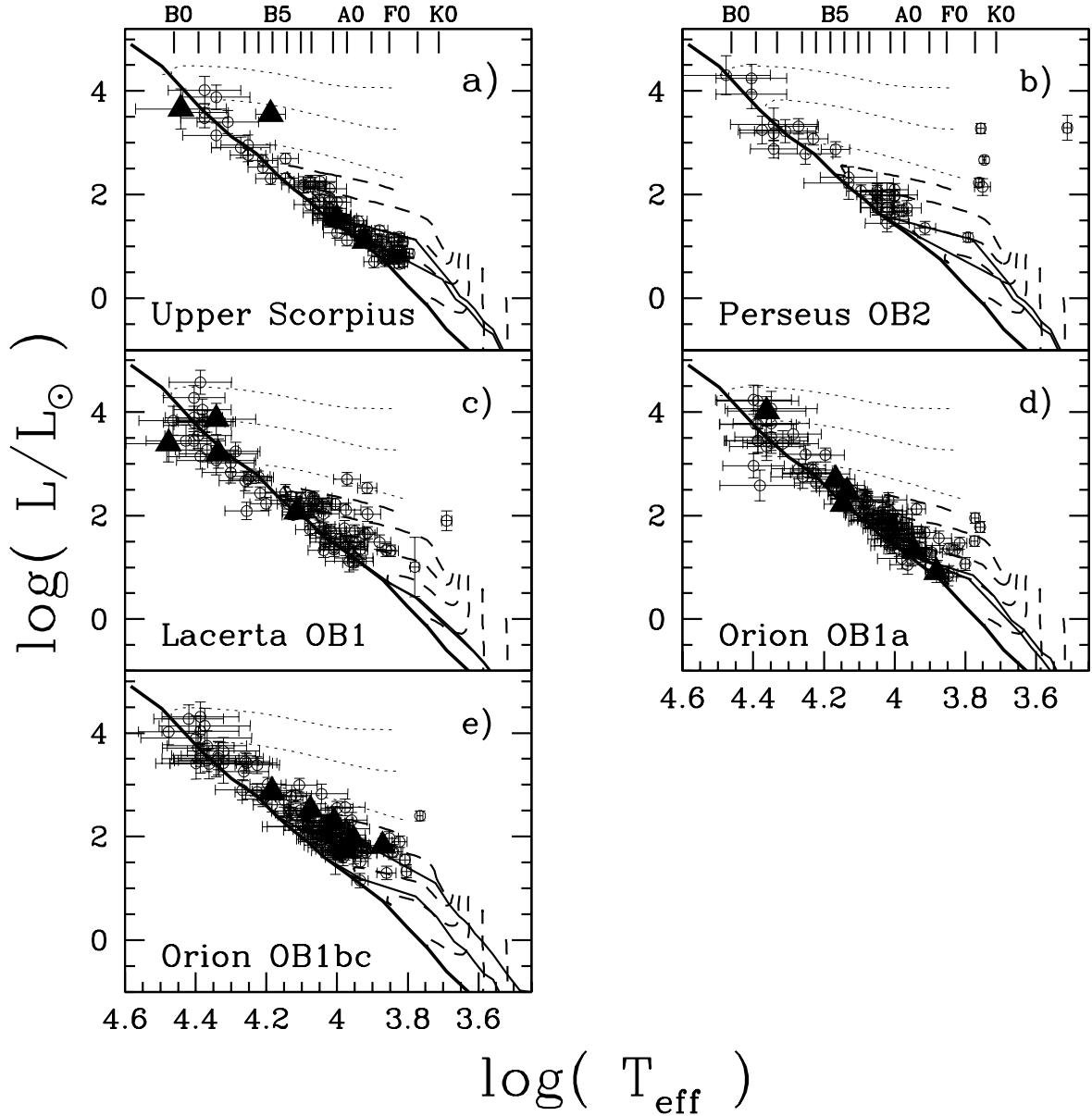}
\caption{HR diagram for the Hipparcos
stars in the OB associations studied in
this work. Emission line stars 
are plotted as solid triangles while open circles indicate  
stars without emission. 
Isochrones from \citet{palla93} for the range of ages of each
association in Table 1 (light solid lines) are shown, as well as
evolutionary tracks from \citet{palla93} 
(dashed lines) corresponding to, from bottom to top, 0.2, 0.6, 1.0, 1.5, 2.0 3.0, and 4.0 \msun;
tracks for 5, 9 and 15 \msun (dotted lines) are from \citet{bernasconi96}.
The ZAMS is represented as a thick solid line. The emission line stars
and the objects without emission share the same region in these plots.}
\label{fig:HRD}
\end{figure}

\begin{figure}
\epsscale{0.7}
\plotone{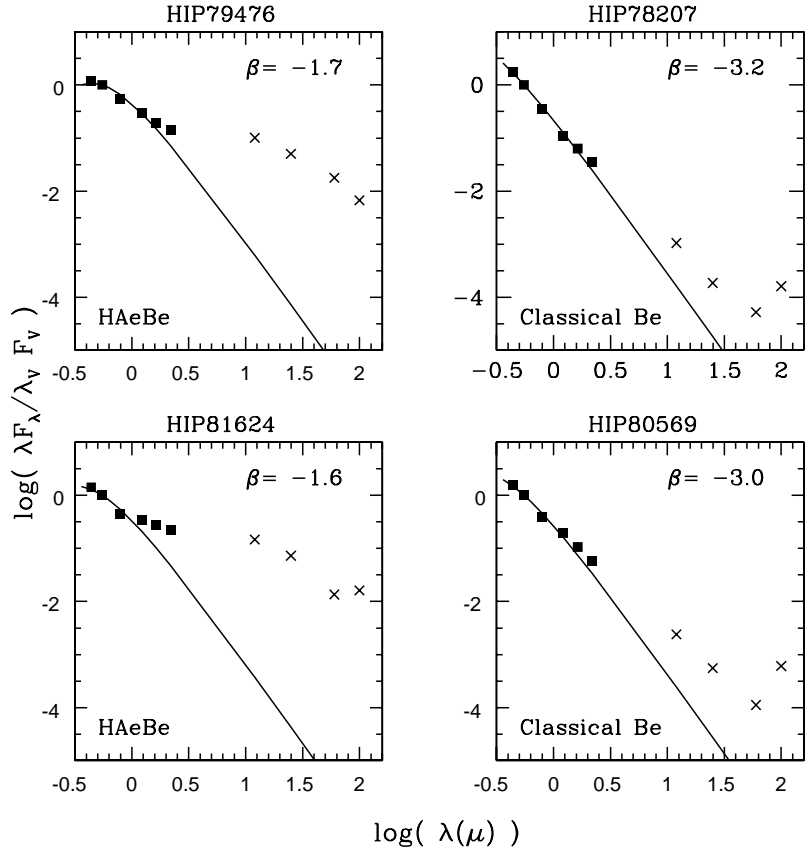}
\caption{SEDs of the emission line stars in the Upper Scorpius OB association. 
Fluxes calculated from magnitudes in 
the B, V, I bands \citep{esa97} and  the J, H, K bands \citep{cutri03}, normalized 
to the flux at V, are plotted as solid squares. The 
IRAS fluxes at 12, 25 60 and 100 $\mu$m are represented by x's.
We show that the $\beta$ index defined by \citet{vieira03}, \S 3.3, can separate the
HAeBe (left column panels) from the CBe (right column panels)}
\label{fig:SED_US}
\end{figure}

\begin{figure}
\epsscale{0.7}
\plotone{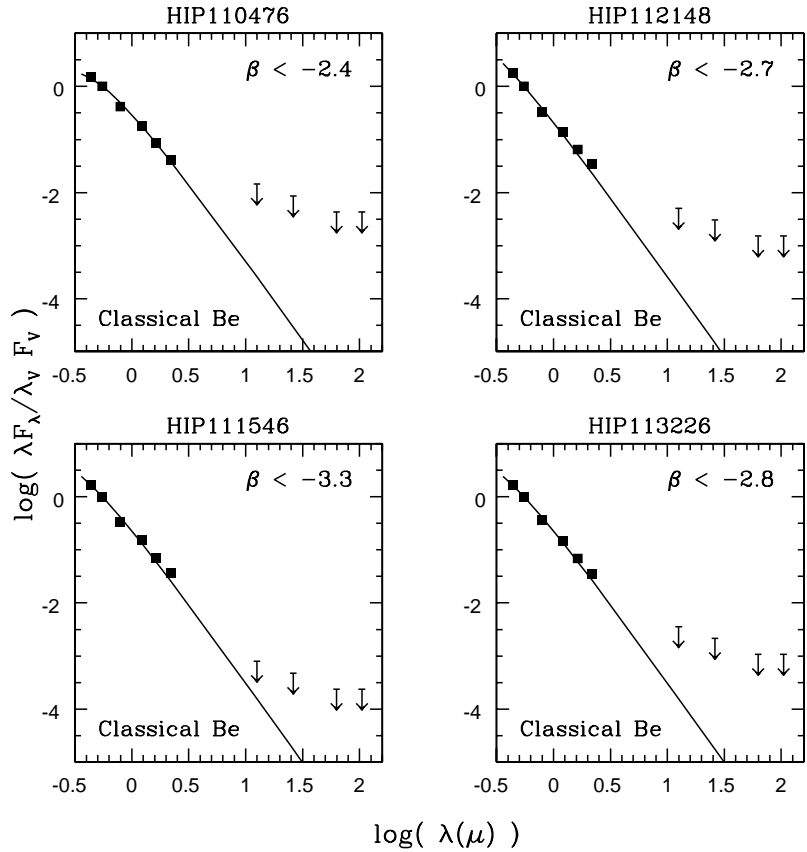}
\caption{SEDs of emission line stars in Lac OB1. Labels as in Figure 5.
The down arrows represent the completeness limit for the IRAS catalog for 
12, 25, 60 and 100 $\mu$m (0.4, 0.5, 0.6 and 1.0 Jy, respectively). The presence of 
inner disks is unlikely in these stars.}
\label{fig:SED_Lac}
\end{figure}

\begin{figure}
\epsscale{0.7}
\plotone{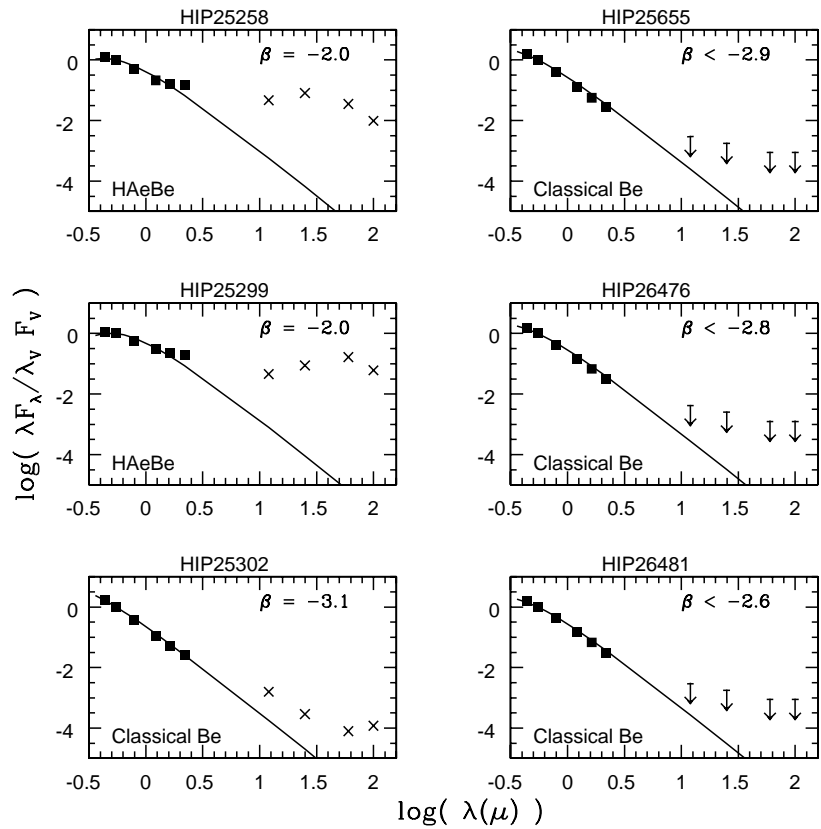}
\caption{SEDs of emission line stars in Ori OB1a. Labels as in Figures 5 and 6.}
\label{fig:SED_OB1a}
\end{figure}

\begin{figure}
\epsscale{0.7}
\plotone{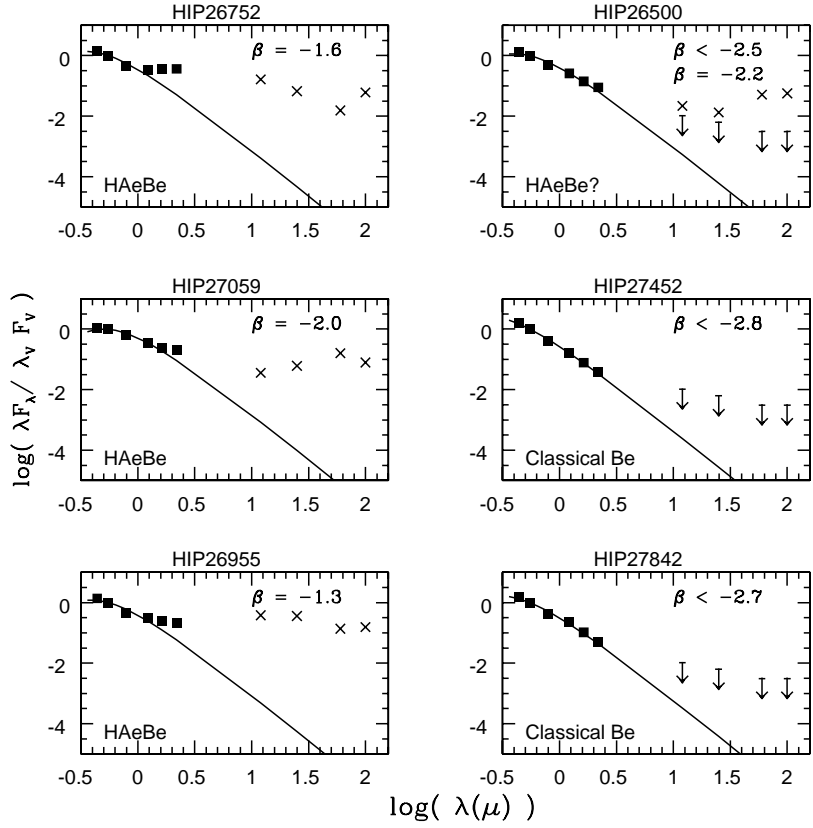}
\caption{SEDs of emission line stars in Ori OB1bc.
Labels as in Figures 5 and 6.
The three objects on the left column panels show clear evidence 
to be HAeBe. 
The status of HIP26500 is uncertain (see \S 4.5).
}
\label{fig:SED_OB1bc}
\end{figure}

\begin{figure}
\epsscale{0.7}
\plotone{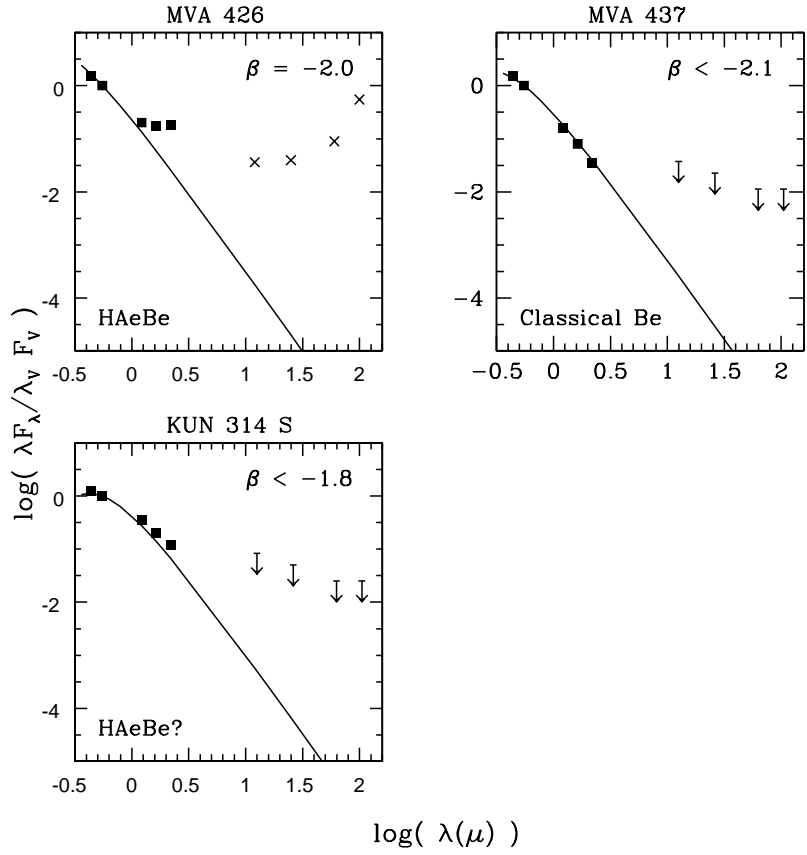}
\caption{SEDs of emission line stars in Trumpler 37. Labels as
in Figures 5 and 6. Only the star MVA 426 has IRAS fluxes
that confirm its classification as HAeBe. The upper limits in the
IRAS fluxes
for MVA 437 suggest that this star is not a HAeBe. The
status of KUN 314S is uncertain (see \S 4.6).
}
\label{fig:SED_Tr37}
\end{figure}

\begin{figure}
\epsscale{0.7}
\plotone{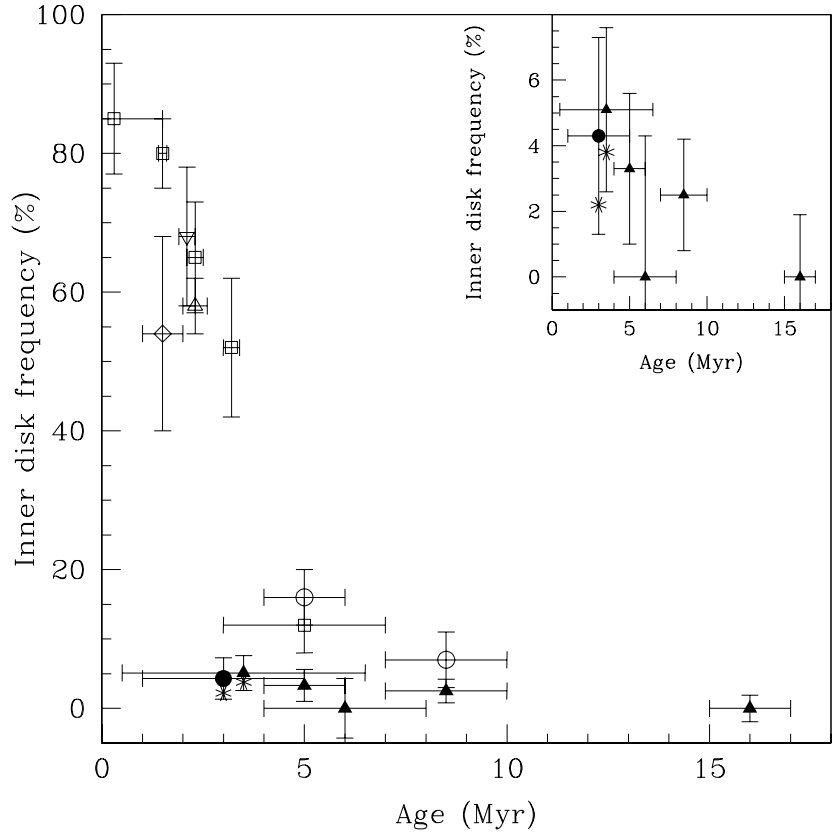}
\caption{Inner Disk frequency as a function of the age of the stellar group.
The solid symbols represent the inner disk frequencies 
for intermediate mass stars estimated from 
the relative numbers of HAeBe stars in each OB association. From left to right, 
the solid triangles represent Ori OB1bc (\S 4.5), Upper Scorpius (\S 4.1), Per OB2 (\S 4.2),
Ori OB1a (\S 4.4), and Lac OB1 (\S 4.3). The solid dot shows the position 
of Tr 37 \citep[\S 4.6; ][]{contreras02}. The asterisks represents the inner 
disk fractions assuming that the stars HIP26500 (Ori OB1bc) and KUN 314S (Tr 37) are 
CBe. 
For comparison, the open symbols represent the disk frequency for the low mass stars,
derived using JHKL observations. From left to right, the open squares 
represent the disks frequency in NGC2024, Trapezium, IC348, NGC2264 and NGC2362 from \citet{haisch01}. The open triangle
and the inverse open triangle show the position on the diagram for Taurus 
\citepalias{kh95}
and Chameleon I \citep{gomez01}. The diamond symbol
represents NGC7129 \citep{gutermuth04}.
The open circles represent our estimates of inner disk frequencies
from the sample of \citet{briceno04} for Ori OB1a and Ori OB1b using the H-K color
(see \S 4.4 and 4.5).
The inner panel shows a zoom of the frequency values 
calculated for the intermediate mass stars. 
}
\label{fig:disk}
\end{figure}

\end{document}